\documentclass[iop,numberedappendix]{emulateapj}

\begin{document}

\font\tenbg=cmmib10 at 10pt
\def \rvecxi{{\hbox{\tenbg\char'030}}}
\def \rvecphi{{\hbox{\tenbg\char'036}}}
\def \rvecdelta {{\hbox {\tenbg\char'016}}}
\def \rvecepsilon {{\hbox {\tenbg\char'017}}}
\def \rvecmu{{\hbox{\tenbg\char'026}}}
\def \rvecOmega {{\hbox {\tenbg\char'012}}}

\title{ON THE ROLE OF FAST MAGNETIC RECONNECTION IN ACCRETING BLACK HOLE SOURCES}

\author{C. B. Singh\altaffilmark{1},
        E.M. de Gouveia Dal Pino\altaffilmark{1},
        and L.H.S. Kadowaki\altaffilmark{1}}

\altaffiltext{1}{Department of Astronomy (IAG-USP),
                 University of Sao Paulo, Brazil;
                 csingh@iag.usp.br; dalpino@iag.usp.br; luis.kadowaki@iag.usp.br}

\begin{abstract}
 We attempt to explain the  observed radio and gamma-ray emission produced in the surrounds of black holes by employing a
magnetically-dominated accretion flow (MDAF) model and fast magnetic reconnection triggered by turbulence.
 In earlier work, standard disk model was used and we refine the model by focussing on the sub-Eddington regime to
 address the fundamental plane of black hole activity.
 The results  do not change substantially with regard to previous work ensuring that the details of the accretion physics are not
 relevant in the magnetic reconnection process occurring in the corona.  Rather our work puts fast magnetic reconnection events as
 a powerful mechanism operating in the core region, near the jet base of black hole sources on more solid ground. For microquasars 
 and low-luminosity active galactic nuclei (LLAGNs) the observed correlation between radio emission and mass of the sources
 can be explained by this process. The corresponding gamma-ray emission also seems to be produced in the same core region.
 On the other hand, the emission from blazars and gamma-ray bursts (GRBs) cannot be correlated to core emission based
 on fast reconnection. 
\end{abstract}


\section{Introduction}
Almost a decade ago \citet[hereafter GL05]{dgdp_lazarian_05} proposed a model for producing jet plasmons and particle
acceleration by magnetic reconnection events in the surrounds of accretion disks around black holes with magnetospheres.
This model predicts that the amount of magnetic power released by reconnection may be more than sufficient to
explain observed flares from black hole mass sources in different scales (from microquasars to LLAGNs)
(see de Gouveia Dal Pino et al. 2010a, hereafter  GPK10; de Gouveia Dal Pino et al. 2010b; Kadowaki et al. 2014, hereafter KGS14).
 Their model invokes the interactions between the field lines anchored onto the black hole horizon and those onto the accretion
 disks around black holes.

Standard disk model \citep{shakura_sunyaev_73} was used in these works to describe the accretion flow around black holes taking into account
near Eddington regime. At high/soft or very high states of the X-ray emission \citep{remillard_mcclintock_06, Fender_etal_04, Fender_etal_09}
the accretion is dominantly via a standard disk extending to the innermost stable circular orbit with a weak corona above the disk. 
However in the low/hard state, it is believed that the accretion flows are geometrically thick, optically thin advection-dominated (ADAFs)
 \citep{NY_95} with an outer geometrically thin, optically thick disk.

Recently \citet{Qiao_Liu_13} considered the cooling of the soft X-ray photons from the underlying accretion disk to the corona, and the 
bremsstrahlung, synchrotron, and  corresponding self-Compton cooling of the corona itself. With the decrease of 
the mass accretion rate, the size of the inner disk decreases and eventually the disk vanishes completely by evaporation. Consequently,
the accretion becomes dominated by an ADAF, in which the X-ray emission is produced by the comptonization of the synchrotron and
 bremsstrahlung photons of the ADAF itself.

The recent simulation work by \citet{Dexter_etal_14} on transient jets in ADAF regime during hard to soft state transition
 has shown that the magnetic reconnection of opposite polarity fields converts magnetic energy into kinetic and thermal energy
 fluxes. The transient jet power depends on the magnetic energy density and time scale over which it is dissipated, not on the
 black hole spin \citep{Dexter_etal_14}.
 Besides, \citet{Sikora_Begelman_13} have suggested that the radio-quiet/loud dichotomy in AGNs could be due to the absence or presence
 of sufficient coherent magnetic fields. To include the dominant role of magnetic fields in the inner region around black holes, 
\citet{Meier_05, Meier_12} proposed that ADAF could be replaced by a magnetically dominated advective flow (MDAF).

GL05 proposed a plausible model in which fast magnetic reconnection episodes in the corona above and below the inner region
 of accretion disks can explain the origin of radio flares observed in the galactic microquasar GRS 1915+105. This might
 happen whenever a large magnetic field arises from the inner accretion disk removing part of the angular momentum so that
 the accretion rate approaches the Eddington rate and pushes the disk magnetic field lines towards the black hole magnetosphere. 
If both magnetic fluxes have opposite polarity then in the presence of anomalously high resistivity or turbulence an event of
 fast magnetic reconnection may take place releasing copious amounts of magnetic power that may accelerate particles to relativistic
 velocities in a first order Fermi process \citep[GL05;][]{Kowal_etal_11, Kowal_etal_12, dgdp_etal_13}.
 The model was further extended to other microquasars, AGNs and young stellar objects (YSOs)  (GPK10). Recently the model was 
applied to a much larger sample of sources than before including LLAGNs, blazars, microquasars and GRBs (KGS14).
They found evidence that the observed correlation between  radio luminosities and the source masses, spanning $10^{10}$ orders of magnitude
 in mass and $10^{6}$ orders of magnitude in luminosity, in microquasars and LLAGNs \citep{Merloni_etal_03, Nagar_etal_05, Fender_etal_04}
 could be naturally explained by this fast magnetic reconnection model. Moreover, they found that the observed gamma-ray emission
 in these sources could be also produced in the same core region. They   also argued that the proposed mechanism could be 
associated with the transition from the low/hard to the high/soft states.

We here revisit a similar scenario related to violent fast reconnection episodes between the field lines of the inner disk corona 
and those that are anchored in the black hole taking into account a sub-Eddington flow. 
We study the same mechanism which comes into play for black hole sources spanning 10 orders of magnitude in mass, but 
instead of assuming as initial state that the accretion disk has already evolved to a Shakura-Sunyaev regime as in the previous works,
 we consider that the system is still in the end of the low/hard state and assume an MDAF disk.
    
In next section, we  present the description of the physical picture regarding the interaction between disk and corona. We also 
present the disk parameters based on an MDAF disk model. Subsequently, the rate of magnetic energy released by fast reconnection is derived.
 In \ref{sec:results} and \ref{sec:conclusions}, we  discuss the results and draw our conclusions. 

\begin{figure}[t]
\centering
\includegraphics[width=0.2\textwidth] {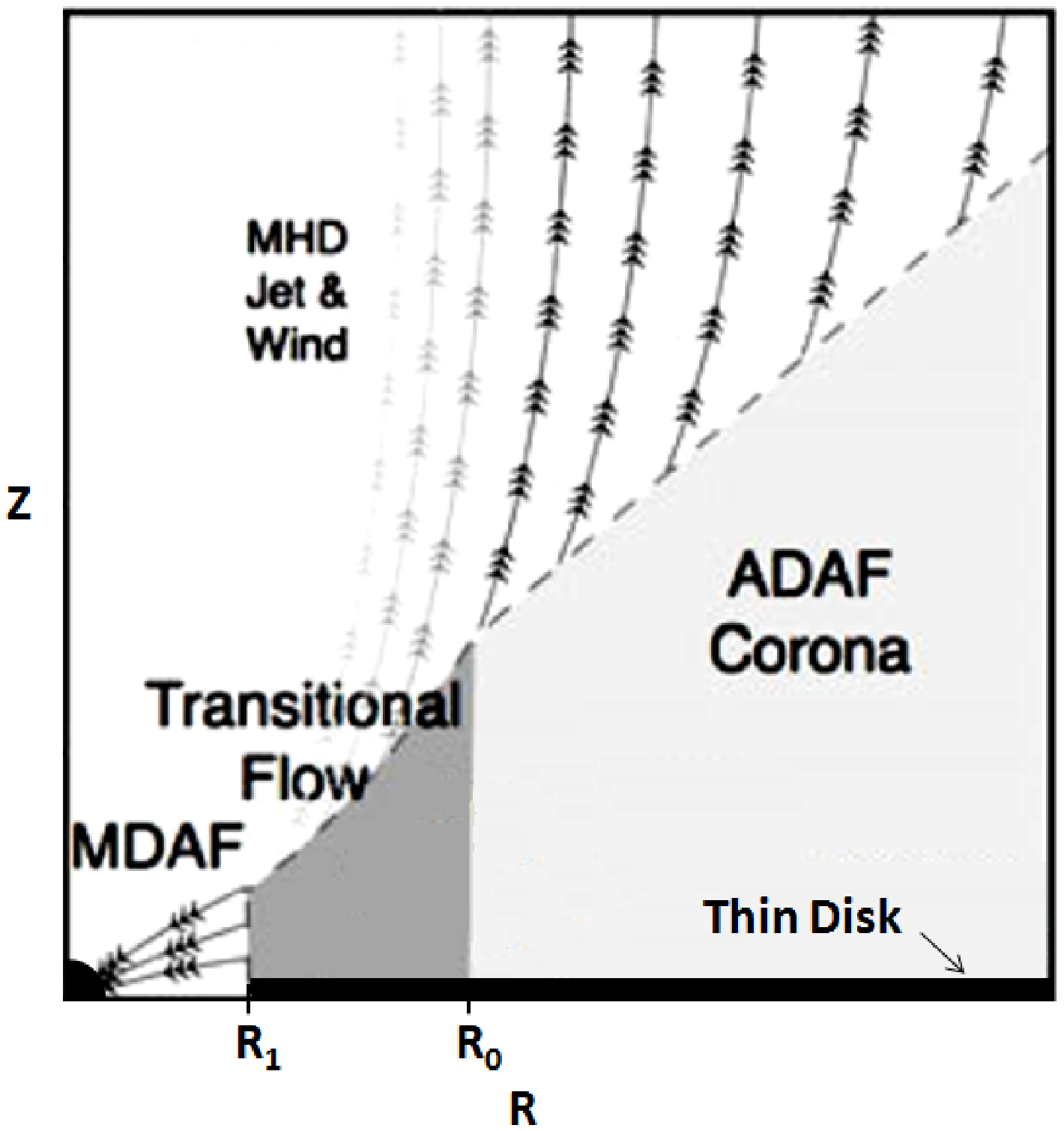}
\includegraphics[width=0.3\textwidth] {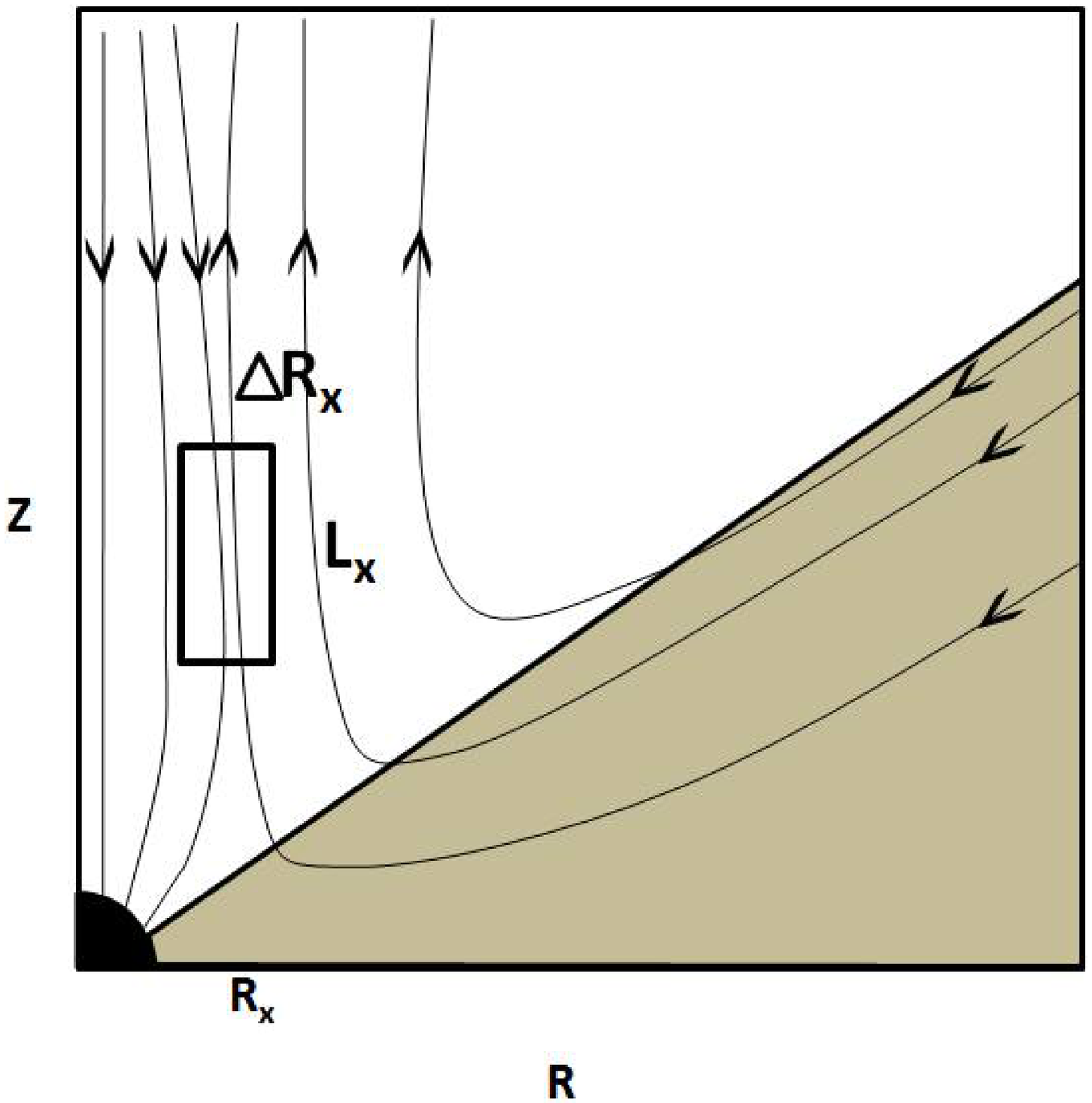}
\vspace{12pt}
\caption{Schematic diagram of our scenario. The upper panel depicts the thin accretion disk truncated at a radius
 $R_{1}$. The transitional flow between MDAF and ADAF  exists in the range $R_1 < R < R_0$ (adapted from \citet{Meier_12}). The lower panel
 depicts our magnetic reconnection scenario in the MDAF region in the surrounds of the black hole ($R_X$ = $R_1$). The magnetic field
 lines anchored into the black hole horizon (not shown in the upper diagram) encounter those rising from the disk, and fast magnetic
 reconnection may occur as described in the text.}
\end{figure}

\section{Revisiting the fast magnetic reconnection scenario considering  MDAF accretion}

\subsection{Inner disk and corona}

Following \citep{Meier_05, Meier_12}, in the disk region $R_{1} < R < R_{0}$, the flow will be in a transitional state between
ADAF and MDAF (Figure 1). In the region $R \le R_{1}$ the flow will be in a truly magnetically-dominated (MDAF) state. Inside $R_{1}$,
 the plasma $\beta$ (= $8\pi p_{g}/B^{2}$) tends to decrease below unity, where $p_{g}$ is the gas pressure and $B$ is the magnetic field
 strength. The location $R_{1}$ is our interest for a fast magnetic reconnection event to happen due to the existence of turbulence in the
 neighbourhood where the transitional flow joins onto the MDAF region (see below).

The masses are scaled in solar units as,
\begin{equation}
M = m M_{\odot},
\end{equation}

and accretion rates in Eddington units,
\begin{equation}
\dot M = \dot m \dot M_{Edd},
\end{equation}

where $\dot M_{Edd} = 1.39 \times 10^{18} m \ {\rm g s^{-1}}$.\\
It is further assumed that, at some radius $R_{0} = 7.3 \times 10^{8}  m \theta^{-1} \ {\rm cm}$, the ion temperature, $T_{i}$ will saturate
 to a finite multiple of $\theta$ of the electron temperature, $T_{e}$
\begin{equation}
\theta = \frac {T_{i}}{T_{e}} \ge 1.
\end{equation}
The value of $\theta$ can lie  between 1 and 820 (Meier 2012). In other words, an MDAF exists somewhere outside the ISCO and
 inside the two-temperature ADAF.\\
Also, as shown in Figure 1, the location of the reconnection region $R_X=R_1$ where $R_{1} = \alpha^{2/3} R_{0}$ is given by
\begin{equation}
R_{X}= R_1= 7.3 \times 10^{8} \alpha^{2/3} m \theta^{-1} \ \rm cm,
\end{equation}
with $\alpha$ being the viscosity parameter.
As in GL05, GPK10 and KGS14, the magnetic field in the inner region can be evaluated from momentum flux balance between the accretion
 flow and the magnetic pressure around the black hole.  In  an MDAF-like corona, this condition implies magnetic fields in the
 Z- and R- directions that depend on $m$, $\dot m$, $\theta$, and $\alpha$, and are expressed as
 \citep{Meier_12},
$$
B_{Z} = 3.34 \times 10^{4} m^{-1/2} \dot m^{1/2} \theta^{5/4} \ {\rm G}, 
$$

$$
B_{R} =  3.84 \times 10^{4} \alpha^{-5/3} m^{-1/2} \dot m^{1/2} \theta^{5/4} \ {\rm G}.
\eqno{(5)}
$$
The corresponding  poloidal field strength is determined as,
$$
B_{p}= 10^{4} (11.15 + 14.89 \alpha^{-10/3})^{1/2} m^{-1/2} \dot m^{1/2} \theta^{5/4} \ {\rm G}.
\eqno{(6)}
$$
Other MDAF coronal parameters such as density, ion temperature and height are respectively given by, 
$$
\rho = 8.9 \times 10^{-10} \alpha^{-2} m^{-1} \dot m \theta^{3/2} \ {\rm g cm^{-3}},
\eqno{(7)}
$$
$$
T_{i} = 10^{9} \theta \ {\rm K},
\eqno{(8)}
$$
$$
H = 6.29 \times 10^{8} \alpha m \theta^{-1} \ {\rm cm}.
\eqno{(9)}
$$

\subsection{Magnetic energy release by fast magnetic reconnection}
Let us assume that the magnetic field anchored into the BH is of the same order of the poloidal magnetic field  in the corona above and below
 the disk in the inner edge of the disk (GL05). This is a reasonable assumption since the BH magnetosphere is built by the dragging of magnetic
 field lines from the accretion disk (\citealt{macdonald_etal_86, neronov_aharonian_07}, GL05). 
Further, to allow reconnection let us assume that the new flux of lines that rise in the MDAF disk corona have opposite  polarity to those
 deposited earlier in the BH magnetosphere (which is possible if dynamo processes occur in the accretion disk; GL05, KGS14
 and references therein).
 As mentioned, in order to extract as much magnetic power as possible to accelerate particles, magnetic 
reconnection has to be fast. The presence of anomalous resistivity \citep{Parker_79, Biskamp_etal_97, Shay_etal_04} or turbulence 
\citep{LV_99, Kowal_etal_09} can speed up the reconnection rate to values near the Alfv\'en velocity ($v_A$).
As described in GL05 and KGS14, the rate of magnetic energy that can be extracted from the 
magnetic contact discontinuity in the corona through reconnection is given by (see rectangular zone in lower panel of Figure 1), 
$$
\dot W = \frac {B_{p}^{2}}{8 \pi} v_{rec} (4 \pi R_{X} L_{X}) = \frac {B_{p}^{2}}{8 \pi} v_{rec} (4 \pi R_{X} A H).
\eqno{(10)} 
$$ 
Here, $0 < A \le 1$.
In the case of fast reconnection driven by turbulence \citep{LV_99} we can derive the magnetic reconnection power in a  
similar way as explained in KGS14. The presence of weak turbulence causes the wandering of the magnetic field lines and induces fast
 reconnection \citep[see also][]{Kowal_etal_09, Kowal_etal_12}.
Assuming that the injection scale of the turbulence ($L_{inj}$) is of the order of the size of the reconnection zone ($L_{X}$),
the reconnection rate is,
$$
v_{rec} \simeq v_{A}M_{A}^{2},
\eqno{(11)} 
$$

$$
v_{A} = \frac {v_{A0}}{(1 + \frac {v_{A0}^{2}}{c^{2}})^{1/2}} = \Gamma v_{A0},
$$

$$
v_{A0} = \frac {B_{p}}{(4 \pi \rho)^{1/2}}.
\eqno{(12)}
$$

\noindent where $M_{A} = v_{inj}/v_{A}$ is the Alfv\'enic Mach number of the turbulence, and $v_{inj}$  the turbulence velocity
 at the injection scale. 
Based on the expression of the turbulence velocity in the transitional layer between ADAF and MDAF, we have \citep{Meier_12}
$$
v_{inj} = 3.7 \times 10^{8} \alpha^{1/2} \theta^{1/2}   \rm cm \, s^{-1}.
\eqno{(13)}
$$
 
Substituting  eqs. (4), (6-9) and (11-13) into equation (10), the magnetic reconnection power released by turbulent
 fast reconnection in the surrounds of the BH is given by (GL05),
$$
\noindent 
\dot W =  3.34 \times 10^{34} \, m \, \dot m  \, \theta \,  A \, \Gamma^{-1} (11.15 \alpha^{10/3} + 14.89)^{1/2} \rm erg \, s^{-1}.
\eqno{(14)}
$$

\noindent We note that  $\dot W \propto m$, which has the same dependence on m as that for standard accretion disk model
 (eq. 15 from KGS14). The dependence with $\dot m$ on the other hand, is stronger in this case ($\dot W \propto \dot m$; while
 in KGS14 it was obtained $\dot W \propto \dot m^{3/4}$). 

\section{Results}
\label{sec:results}
As stressed in KGS14, the magnetic reconnection power that is produced in an anomalous resistivity model does not cover most of
 the observed radio and gamma-ray emissions of the sources, so in what follows we consider the turbulent driven fast reconnection model only.
Besides, the MDAF scenario naturally drives turbulence in the corona, as remarked.

Figure 2 compares the calculated fast magnetic reconnection power driven by turbulence as derived in this work
\footnote{We note that the adopted range of values for $\theta$ and $A$  ensure the
nearly collisionality condition required by the MHD equations and the fast reconnection driven by turbulence (see details in KGS14).
 With this parametric space $L_X= A H$ spans between $0.02 R_S$ and $ \sim 44 R_S$, and $R_X$ between
 $1.1 R_S$ and $\sim 186 R_S$,} where $R_S =  2.95 \times 10^{5} m$  cm is the Schwarzschild radius. 
(range of free parameters $1 \le m \le 10^{10}$; $5 \times 10^{-4} \le \dot m \le 0.05$;  $0.003 \le \alpha \le 0.3$,
 $0.01 \le A \le 1$; $1 \le \theta \le 46$) with the calculated power in KGS14 employing the standard accretion model rather than MDAF.
 We find that, in spite of the inherent differences in the assumptions and parametrization
(the parametric space in KGS14 spans  $R_{X}/R_{S} =$ 6$; 1 \le m \le 10^{10}$;
$5 \times 10^{-4} \le \dot m \le 1$;  $0.05 \le \alpha \le 0.5$ and 
$0.06 R_{S} < L_{X} < 17.5 R_{S}$), both models produce very similar ranges of values for the magnetic reconnection power,
 therefore confirming the earlier prediction in
 GPK10 that the details of the accretion model should not affect much the results regarding the fast reconnection power extracted from the
 coronal regions around  BHs. The models are also compared with the observed correlations between the core radio luminosity and
 the BH mass found for microquasars and LLAGNs by \citet{Nagar_etal_02, Nagar_etal_05} and \citet{Merloni_etal_03}.
  We see that the slope dependence of the magnetic power released by turbulent reconnection with the source mass is very similar
 to the observed radio luminosity-source mass correlations for these sources. 

\begin{figure*}[t]
\centering
\includegraphics[width=0.4\textwidth] {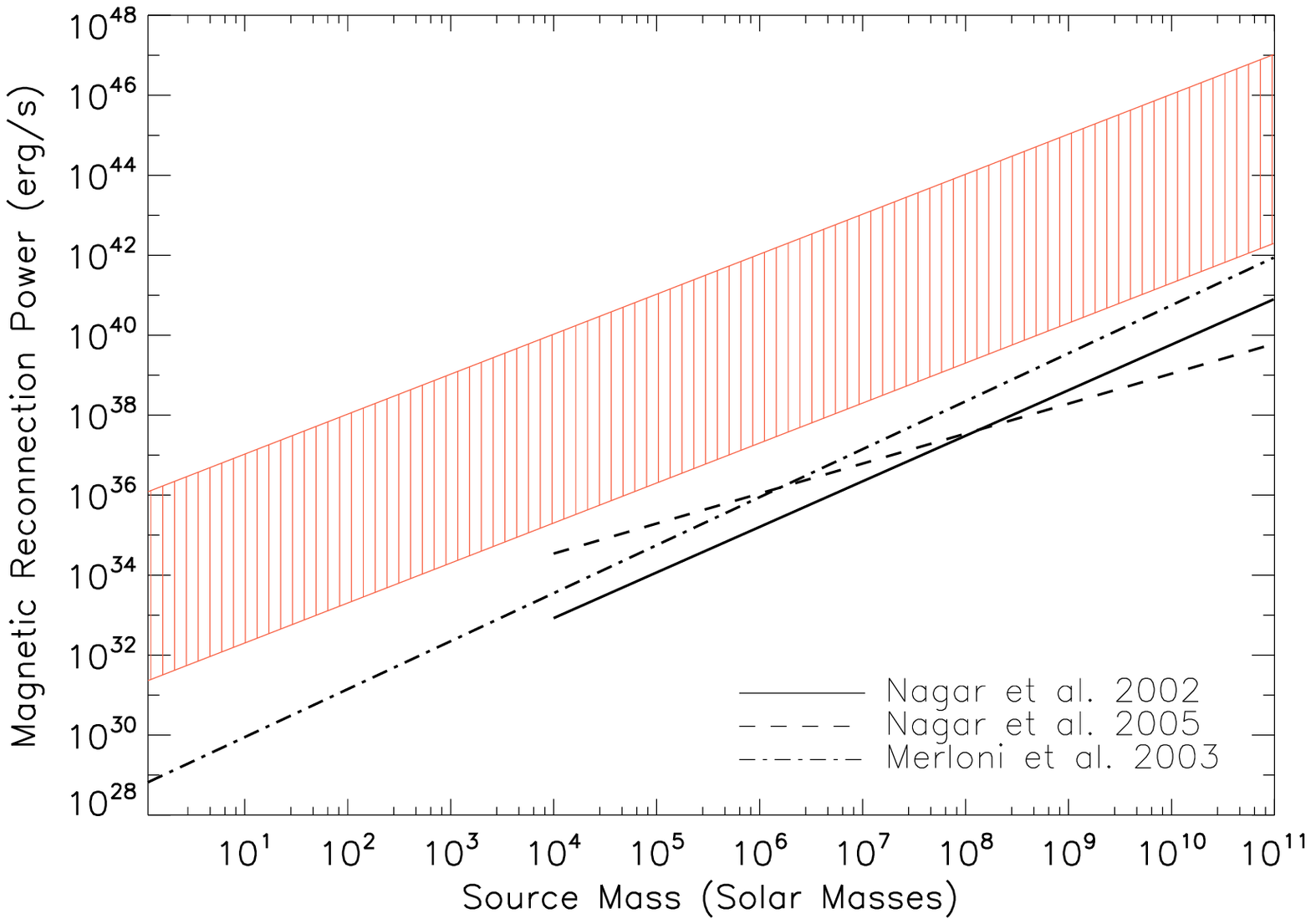}
\includegraphics[width=0.4\textwidth] {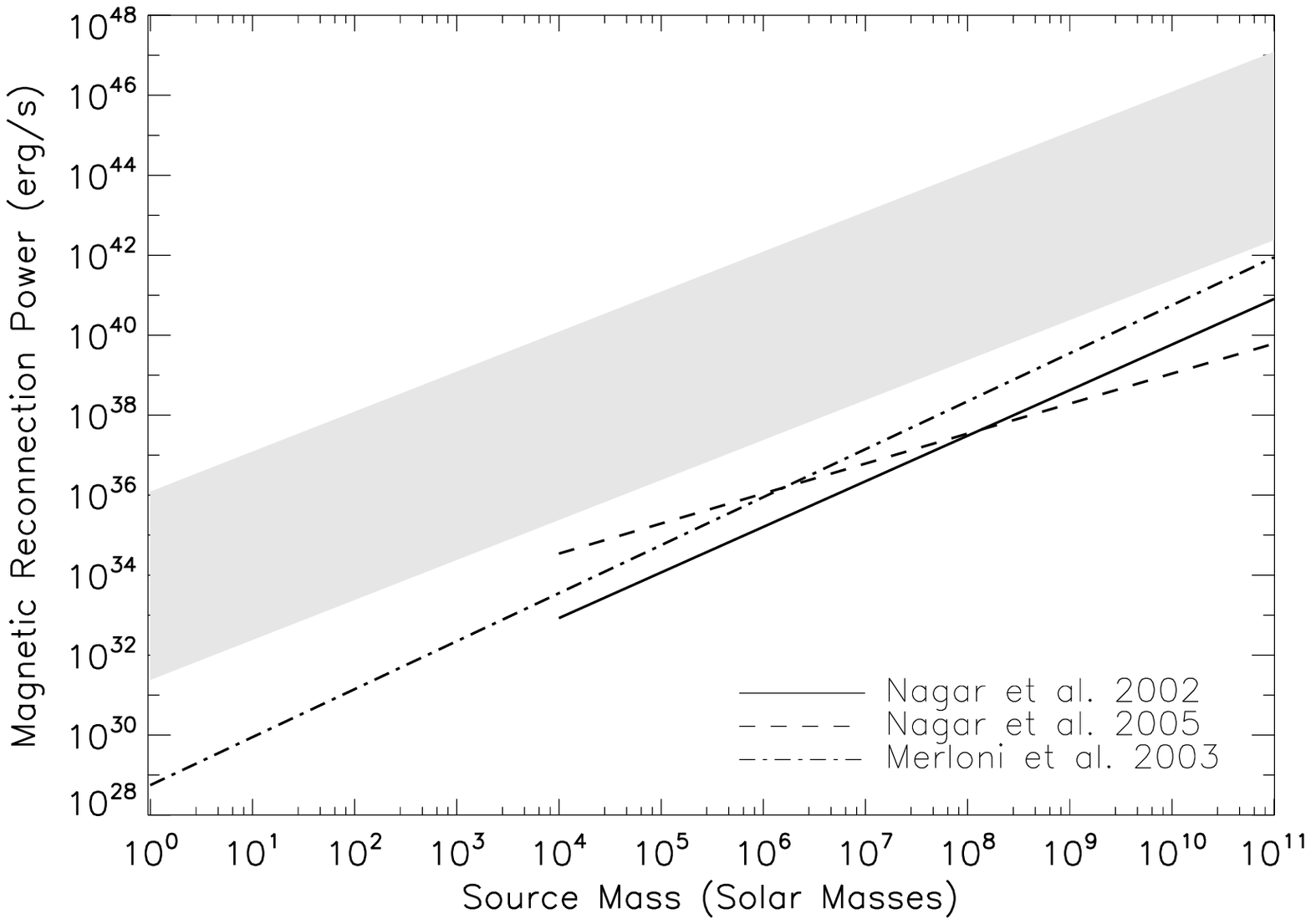}
\caption{Turbulent driven magnetic power against BH source mass calculated in this work (eq. 14) (right panel) compared to the one
 calculated in KGS14 (left panel; see text for details). In both panels, the continuous and dashed black lines correspond to the
 observed correlations between the BH mass and the core radio luminosity found for LLAGNs by \citet{Nagar_etal_02} and \citet{Nagar_etal_05}
 respectively; and the dot-dashed  line corresponds to observed correlations for AGNs and microquasars by \citet{Merloni_etal_03}.
}
\end{figure*}

\begin{figure*}[t]
 \centering
 \includegraphics[width=0.5\textwidth] {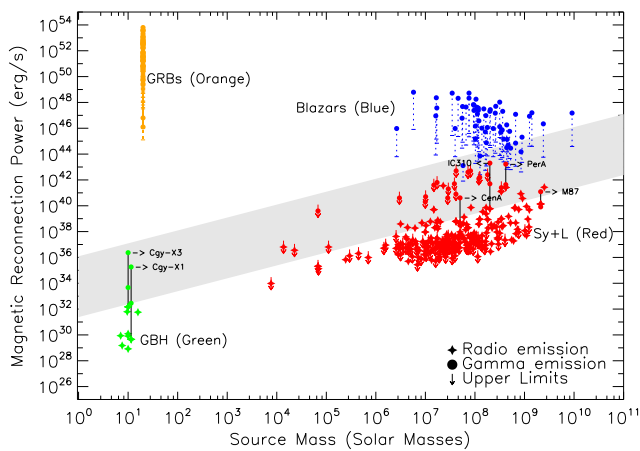}
\caption{ 
Turbulent driven magnetic power (eq. 14) against BH source mass compared to the observed  emission of LLAGNs (LINERS and Seyfert galaxies),
  microquasars (GBHs), blazars and GRBs. The core radio emission of the GBHs and LLAGNs is represented by red and green diamonds, while
 the gamma-ray emission of these two classes is represented by red and green circles, respectively.  In a few cases for which there is observed
 gamma-ray luminosity from MeV/GeV to TeV ranges, it is plotted the maximum and minimum values linking both circles with a vertical black
 line that extends down to the radio emission of each  of these sources. The inverted arrows associated to some sources indicate that the
 gamma-ray emission is an upper limit only.  For blazars and GRBs only the gamma-ray emission is depicted, represented in blue and orange
 circles, respectively. The vertical dashed lines correct the observed emission by  Doppler boosting effects (see more details in KGS14).
}
\end{figure*}

As in KGS14, in Figure 3 we compare the calculated fast magnetic reconnection power driven by turbulence (eq. 14)  with the observed nuclear
 radio and gamma-ray luminosities of a large sample (of more than 270 sources) including microquasars (or galactic black hole binaries - GBHs),
 LLAGNs, blazars, and gamma-ray bursts (GRBs) (see KGS14 and references therein for a detailed description of
 this sample). The radio emission is represented by diamond symbols, with red and green colours representing LLAGNs and microquasars, respectively.
The observed radio emissions of a few microquasars (Cyg X-1 and Cyg X-3) and LLAGNs (Cen A, IC 310, Per A and M87) which have been extensively
 studied in multi-wavelength campaigns are highlighted in the figure. The  released fast magnetic reconnection power driven by turbulence,
 corresponding to  fiducial parametric space, largely exceeds the observed radio luminosities. The figure indicates that only a small fraction 
 of this  power would be enough to explain the observed radio emission for most of the  sources.
 
As discussed in KGS14, the accelerated relativistic electrons and protons are likely to cool via other mechanisms (inverse Compton,
 Synchrotron self Compton, proton-proton and proton-photon processes) that may eventually lead to  very high energy emissions. Thus we have
 included in Figure 3 also the observed gamma-ray luminosity, which is  available for only 23 of the LLAGNs and microquasars
 sources in the sample. This is represented by red and green  circle symbols, respectively. These circles correspond to the emission
 of several Seyfert galaxies (most of which show only upper limits in the GeV band), four radio galaxies and two microquasars (Cyg X-1 
and Cyg X-3). The calculated magnetic energy power  appears to be sufficient enough to produce also the gamma-ray emission, however
 what can be seen with certainty is that this emission follows the same trend as that of the radio emission and that both seem to be
 correlated. This may imply that both gamma-ray and radio emissions are likely to be produced in the same core region around the black hole
 and are powered by the same acceleration mechanism. 
 Figure 3 also depicts  gamma-ray emission of blazars (represented by blue circles).  This sample contains 32 
blazars studied \citep{nemmen_etal_12}.
In spite of the Doppler correction, the blazars emission does not appear to follow the same trend
 as that of LLAGNs and microquasars. Their radio emission (not shown here, but in KGS14) does not follow the trend either.
This is likely because the emission in blazars may originate from further out than the core region  
and may be produced by another population of relativistic particles probably at the shock along the inner jet. This is
 consistent with the fact that the jet in blazars  points towards the line of sight screening most of the inner core radiation.
 
Further, Figure 3 also includes gamma-ray emission of 54 GRBs \citep{nemmen_etal_12} which do not appear to follow any trend,
 similar to blazars. It can be suggested that the gamma-ray emission in case of both blazars and 
GRBs cannot be associated with the emission process resulting from fast magnetic reconnection events in the core region;
it is likely to be predominantly coming from the optically thin jet (KGS14).

\section{Discussion and concluding remarks}
\label{sec:conclusions}
We have extended the earlier works by GL05, GPK10 and KGS14 investigating the conditions for fast reconnection between the
 magnetic field lines that rise from the accretion disk and the lines anchored into the black hole (BH) horizon.
 Distinctly from these previous works which adopted the standard thin, optically thick model of disk accretion, in this  study
 we considered MDAF accretion in the inner region around the BH, which includes the dominant role of the magnetic field and
 is suitable for sub-Eddington sources \citep{Meier_05, Meier_12}. This model produces results which are similar to those
 obtained in the earlier works.
 The calculated magnetic power released by fast reconnection driven by turbulence versus source mass is consistent with 
the observed correlation between radio emission and source mass for microquasars and LLAGNs. This may be useful for interpretation
 of the fundamental plane of black hole activity which correlates the radio and x-ray emission of these sources with their black hole mass
\citep{Merloni_etal_03, Falcke_etal_04, Huang_etal_14}. The investigation of the X-ray
emission, which is directly related with emission processes in the accretion disk, is out of the scope of this work which focussed
 on coronal emission. Nevertheless, our model suggests a simple  interpretation for the existence of these empirical correlations.
 
 As argued in GPK10 and KGS14, since  fast magnetic reconnection and the associated emission flares are strongly dissipative phenomena
 that lead to  partial destruction of the equilibrium configuration in the inner accretion disk/corona region, this mechanism could be
 related to the transition from the hard or low/hard to the high/soft X-ray state \citep[see also][]{Huang_etal_14}.
 
The observed gamma-ray emission of these sources is also well correlated with the radio luminosity, and the calculated fast magnetic 
reconnection power  is large enough to produce them both in the core region (Figure 3). Furthermore, as mentioned in KGS14, the results 
here do indicate that the fast reconnection is relatively insensitive to the accretion disk model. The similarity between the results
 here produced with MDAF accretion and those of KGS14 with standard disk accretion is striking (Figure 2). 
The only difference is that in order to magnetic reconnection to produce both emissions, the sources require in general accretion rates 
$5 \times 10^{-4} < \dot m \le 0.05$ in the MDAF case and  $\dot m > 0.05$ in the standard accretion case (KGS14).
 This difference is due to the inherent physical assumptions of each model as described. 
 
  Finally, our work further supports the observed correlation between GRBs and blazars \citep{nemmen_etal_12} and suggests that
 the gamma-ray and radio emission from such sources cannot be produced by fast magnetic reconnection in the core region. It is
  rather originated further out in the jet. 
  
To conclude, the results above mainly connect the radio and gamma-ray emission from low luminosity compact sources to  magnetically
 dominated reconnection process in the nuclear region of these sources whether the accretion flow model is standard disk-corona
 (KGS14) or MDAF (this work). As further step, more general analytical   and numerical studies will be needed to explore the scenario here 
presented more realistically. Also,  the reproduction of observed non-thermal spectral energy distributions of different sources employing
 the acceleration mechanism above will help to further test it \citep[see attempts in this direction in][]{khiali_etal_14}.  
 
\acknowledgments
 This work has been partially supported by the Brazilian agencies FAPESP ($2013/09065-8$, $2011/51275-4$), and CNPq 
($300083/94-7$, $142220/2013-2$).

\end{document}